\documentclass[twocolumn, switch]{article}
\usepackage{preprint}
\usepackage{hyperref}
\usepackage[numbers,square]{natbib}

\usepackage{booktabs}
\usepackage{graphicx}
\hypersetup{colorlinks=true, linkcolor=purple, urlcolor=blue, citecolor=cyan, anchorcolor=black}

\usepackage[utf8]{inputenc}	% allow utf-8 input
\usepackage[T1]{fontenc}
\usepackage{xcolor}
\usepackage{lineno}					% Line numbers
\usepackage{tikz} 					% ORCiD insertion

\usepackage{newfloat}
\DeclareFloatingEnvironment[name={Supplementary Figure}]{suppfigure}
\usepackage{sidecap}
\sidecaptionvpos{figure}{c}

\usepackage{titlesec}
\titlespacing\section{0pt}{12pt plus 3pt minus 3pt}{1pt plus 1pt minus 1pt}
\titlespacing\subsection{0pt}{10pt plus 3pt minus 3pt}{1pt plus 1pt minus 1pt}
\titlespacing\subsubsection{0pt}{8pt plus 3pt minus 3pt}{1pt plus 1pt minus 1pt}

\definecolor{lime}{HTML}{A6CE39}
\DeclareRobustCommand{\orcidicon}{
	\begin{tikzpicture}
	\draw[lime, fill=lime] (0,0)
	circle [radius=0.16]
	node[white] {{\fontfamily{qag}\selectfont \tiny ID}};
	\draw[white, fill=white] (-0.0625,0.095)
	circle [radius=0.007];
	\end{tikzpicture}
	\hspace{-2mm}
}

\title{Do Small Language Models Know When They’re Wrong? Confidence-Based Cascade Scoring for Educational Assessment}

\usepackage{authblk}
\usepackage{float}
\author[1\thanks{\texttt{tylerb@khanacademy.org}}]{Tyler Burleigh, PhD\href{https://orcid.org/0000-0002-9064-8876}{\orcidicon}}
\affil[1]{Khan Academy}

\begin{document}

\twocolumn[\begin{@twocolumnfalse}

\maketitle

\begin{abstract}
Automated scoring of student work at scale requires balancing accuracy against cost and latency. In ``cascade'' systems, small language models (LMs) handle easier scoring tasks while escalating harder ones to larger LMs --- but the challenge is determining which cases to escalate. We explore verbalized confidence --- asking the LM to state a numerical confidence alongside its prediction --- as a routing signal. Using 2,100 expert-scored decisions from student-AI math conversations, we evaluate cascade systems built from GPT-5.4, Claude 4.5+, and Gemini 3.1 model pairs. We find that: (1) confidence discrimination varies widely across small LMs, with the best achieving AUROC 0.857 and the worst producing a near-degenerate confidence distribution; (2) confidence tracks human scoring difficulty, with lower LM confidence where annotators disagreed and took longer to score; (3) the best cascade approached large-LM accuracy (kappa 0.802 vs. 0.819) at 76\% lower cost and 61\% lower latency. Confidence discrimination is the bottleneck: the two small LMs with meaningful confidence variance yielded cascades with no statistically detectable kappa loss, while the third --- whose confidence was near-degenerate --- could not close the accuracy gap regardless of threshold. Small LMs with strong discrimination let practitioners trade cost for accuracy along the frontier; those without it do not.\\
\end{abstract}

\keywords{automated scoring, language models, confidence calibration, cascade systems, educational assessment, conversation-based assessment}

\vspace{0.5cm}

\end{@twocolumnfalse}]

%%%%%%%%%%%%%%%  Main text   %%%%%%%%%%%%%%%

\section{Introduction}

Reliable automated scoring is critical for large-scale educational platforms. Language models (LMs) can score student work with near-human accuracy --- prompted with a rubric and student response, they return a judgment without task-specific fine-tuning --- but present a tradeoff: larger LMs tend to score more accurately, but cost more and take longer to respond. The cost gap can be substantial: at the time of writing, GPT-5.4 costs roughly 10 times what GPT-5.4-nano does per token \citep{openai2026pricing}. Latency matters just as much in synchronous applications: in conversation-based assessments, the system evaluates student responses at each turn; in computerized adaptive tests, item selection depends on scorer results. Delays from larger LMs in either case can frustrate students.

Cascade scoring systems address this tradeoff by giving a small LM the first attempt at every scoring decision. When the small LM is accurate, the process stops there, saving the cost and latency of a larger LM. But some decisions are harder, and the small LM is more likely to get them wrong; the system must identify and escalate these. The question is how to identify them.

In this study, we investigate verbalized confidence --- asking the LM to state a numerical confidence alongside its prediction --- as a routing signal. When a small LM reports lower confidence, that decision is routed to a larger LM. For this to work, confidence must separate accurate from inaccurate predictions --- a property called \textit{discrimination}. Without that, the cascade has no basis for routing and offers no advantage over simply using the large LM for everything. Recent work documents improved discrimination on reasoning and knowledge tasks with newer LMs \citep{xiong2024confidence, tao2025revisiting, tao2025human}, but to our knowledge, no prior work has tested it in educational scoring.

But good discrimination alone is not enough. Even if confidence reliably separates accurate from inaccurate predictions, the cascade only pays off under two further conditions: the large LM must be meaningfully more accurate than the small one on escalated cases, and the small LM must handle enough decisions to offset the cost of running both.

Before trusting confidence as a routing signal, we also need evidence that it tracks task difficulty rather than producing arbitrary numbers. If LMs are less confident on the same decisions that were harder for human scorers, that is at least consistent with the signal reflecting genuine ambiguity, though, as we discuss later, correlation with difficulty proxies does not establish the mechanism.

\subsection{Related work}

Recent work has shown that off-the-shelf LMs, prompted with a rubric, can score constructed responses with near-human accuracy across a range of items and subjects \citep{frohn2025scoring}. The present study builds on that premise and asks how to make such scoring more efficient through confidence-based routing.

Educational measurement has a long history of routing uncertain automated scores to more reliable scorers. E-rater withholds scores for responses that advisory flags identify as unscorable and uses discrepancy thresholds to route disagreements to a second human rater \citep{burstein2013erater, ramineni2013guidelines}. The Intelligent Essay Assessor evaluates ``the confidence with which it can score [a response] accurately'' and directs low-confidence cases to human raters \citep{foltz2013iea}. Later work explored confidence-based routing in short-answer scoring, where routing low-confidence predictions to human raters can reduce serious scoring errors to zero on roughly half the test data \citep{funayama2020cse, funayama2022hitl}.

In a related line of work, \citet{xiao2025dualprocess} use a fine-tuned LLaMA3-8B that scores essays quickly in a fast mode and routes low-confidence cases to its own slower, explanation-generating mode. Our study extends this line of work in two ways. First, we use verbalized confidence --- the LM states a numerical confidence alongside its prediction --- rather than internal model signals. Second, we escalate to a larger, independent LM rather than a slower pathway within the same model.

Methods for estimating the confidence of LM decisions include inspecting token-level probabilities and measuring response consistency across repeated samples \citep{geng2024survey, tao2025human, xiong2024confidence}. Both have practical drawbacks for our setting: commercial LM providers often restrict access to token probabilities; the Anthropic API does not expose them at all, and OpenAI disables them for reasoning models like GPT-5.4; repeated sampling (e.g., the N=10 sampling approach of \citet{xiong2024confidence}) multiplies cost, which defeats the purpose of a cascade designed to reduce it. A third alternative, verbalized confidence, avoids both issues: the LM simply expresses a numerical confidence together with its decision. This approach has gained traction for its ease of implementation. Recent work suggests it is increasingly competitive with token-level methods, though this was not always the case.

Early results on confidence discrimination via verbalization were modest: \citet{xiong2024confidence} report AUROC (where 0.5 is chance) below 0.63 across tasks ranging from object counting to answering professional law queries, though scores improve considerably from GPT-3 to GPT-4 (0.513 to 0.627). With more recent LMs, \citet{tao2025revisiting} find that GPT-4.1, Grok 3, and Qwen3-235B exceed 0.7 across reasoning and knowledge tasks drawn from MMLU-Pro, and \citet{tao2025human} report similar results with GPT-5 Mini on question-answer benchmarks like SimpleQA and PopQA.

Beyond discrimination, LM cascades have been explored for cost reduction on general reasoning and knowledge tasks. \citet{chen2023frugalgpt} show that chaining LM APIs can reduce cost on classification and QA tasks. \citet{chuang2025routing} benchmark eight uncertainty methods for routing across 1,500+ configurations and find that trained probes and perplexity outperform verbalization, but their small LMs ranged from 1--8B parameters, well below the frontier small models we test.

Our study tests verbalized confidence in rubric-based educational scoring --- to our knowledge, the first exploration of LM-to-LM verbalized confidence routing in this domain, and examines whether the signal tracks human-judged difficulty, not just whether it improves aggregate accuracy.

\subsection{Research questions}

\begin{itemize}
\item \textbf{RQ1:} Does verbalized confidence discriminate between accurate and inaccurate scores well enough to serve as a routing signal?
\item \textbf{RQ2:} Does AI confidence track human scoring difficulty?
\item \textbf{RQ3:} Can a confidence-routed cascade approach large-LM accuracy while reducing cost and latency?
\end{itemize}

\section{Methods}

\subsection{Assessment context}

The scoring decisions in this study come from Explain Your Thinking (EYT) items, a conversation-based assessment (CBA) format used on Khan Academy's high-school mathematics interim assessments \citep{burleigh2025prepilot}. In CBA, an AI chatbot conducts a multi-turn dialogue with students, using adaptive follow-up questions to probe conceptual understanding \citep{yildirim2023cba}. Each EYT item has two parts. In Part 1, students answer a mathematics problem (e.g., multiple-choice or numeric entry). In Part 2, the AI engages the student in a follow-up conversation about their reasoning. The conversation begins with a predefined prompt (e.g., ``Look back at your work. Does Julia's first step make sense to you?''), then the AI generates questions tailored to the student's responses. The dialogue ends when the student satisfies all rubric criteria, reaches a four-turn limit, or exits early.

In the production system, separate modules handle conversation generation and scoring; the LMs evaluated in this study simulate only the scoring module, and all conversations were collected before scoring began. Each item has multiple scoring criteria representing key aspects of conceptual understanding. A criterion definition includes an indicator statement, evaluation notes, and examples of correct and incomplete student work. Each criterion is evaluated independently as a binary judgment (met / not met) based on the full conversation --- both by human raters and by the LM --- so a conversation with two criteria produces two separate scoring decisions. This criterion-level decision is the unit of analysis throughout the study.

\subsection{Dataset}

\begin{table}[H]
\centering
\textbf{Table 1.} Dataset characteristics.

\smallskip
\begin{tabular}{ll}
\toprule
Metric & Value \\
\hline
Items & 4 \\
Conversations per item & 300 \\
Included criteria & 7 \\
Scoring decisions & 2,100 \\
Annotators per decision & 3 \\
Annotation rows & 6,300 \\
Class balance (correct) & 25\% \\
Class balance (incorrect) & 75\% \\
\bottomrule
\end{tabular}
\end{table}

We draw from 1,200 EYT conversations across 4 high-school math items (2 algebra, 2 geometry, 300 conversations each), with 1--2 criteria per item, producing 2,100 scoring decisions in total. Each decision is independently rated by 3 expert annotators.

All criteria have Fleiss' kappa \citep{fleiss1971measuring} at or above 0.70 (``substantial'' agreement per \citet{landis1977measurement}), ensuring reliable ground truth for evaluating AI confidence.

\begin{table}[H]
\centering
\textbf{Table 2.} Human inter-rater agreement by item and criterion.

\smallskip
\begin{tabular}{lll}
\toprule
Standard & Criterion & Human kappa \\
\hline
G-SRT.C.6 & 1 & 0.802 \\
F-IF.A.1 & 1 & 0.858 \\
F-IF.A.1 & 2 & 0.713 \\
G-SRT.A.2 & 1 & 0.800 \\
G-SRT.A.2 & 2 & 0.910 \\
F-IF.B.6 & 1 & 0.797 \\
F-IF.B.6 & 2 & 0.749 \\
\bottomrule
\end{tabular}
\end{table}

The triple-annotation design serves a dual purpose. It provides ground truth via majority vote --- we define the ground-truth label for each scoring decision as the judgment given by at least 2 of 3 annotators --- and, crucially, it provides a measure of human scoring difficulty: patterns of agreement and disagreement among annotators serve as a proxy for the inherent ambiguity of each scoring decision.

To illustrate what a scoring decision looks like in practice, the following example shows the three inputs the LM receives: the math problem from Part 1, one rubric criterion, and the student-AI conversation from Part 2.

\includegraphics[width=0.7\linewidth]{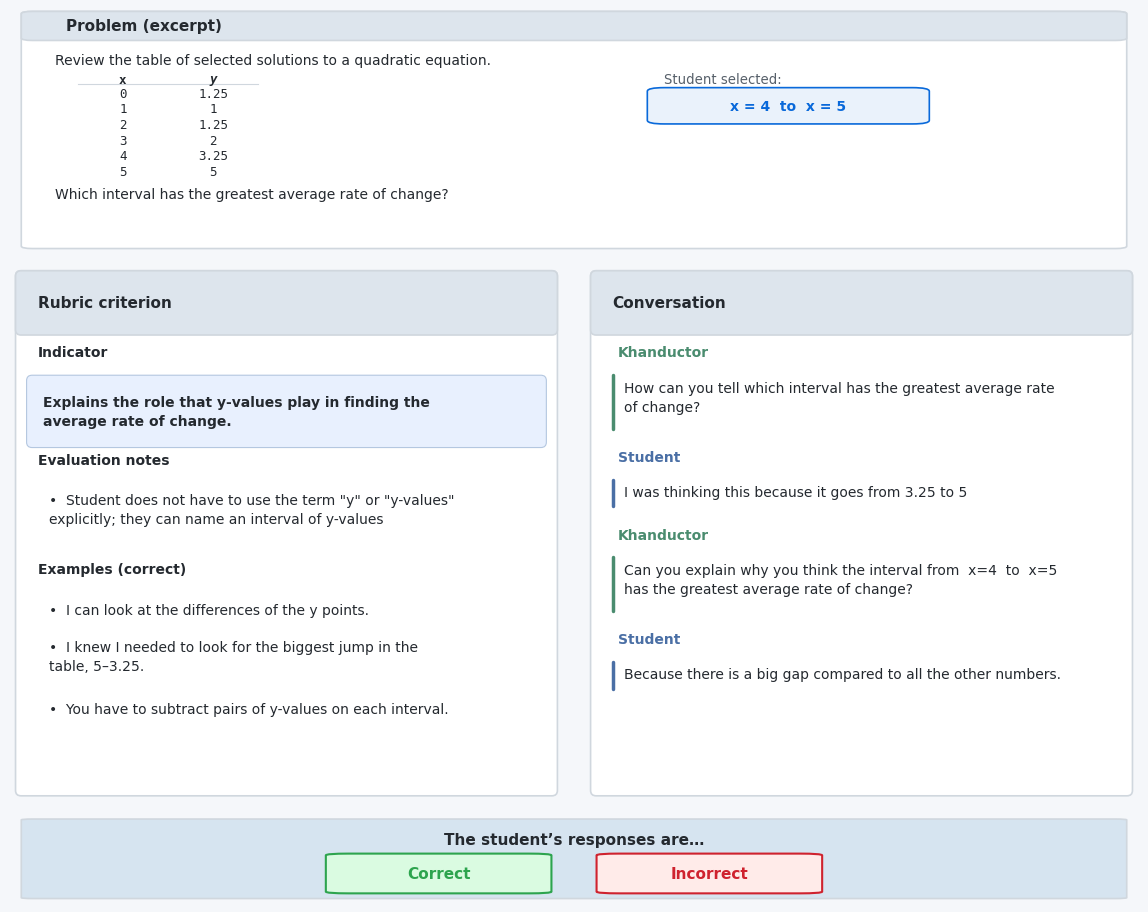}

\textbf{Figure 1.} The scoring task for item F-IF.B.6 (algebra). The annotator sees the math problem, rubric criterion with evaluation guidance, and the student-AI conversation, then judges whether the student's responses satisfy the criterion.

\subsection{Human uncertainty measures}

We define two complementary proxies for human scoring uncertainty, which serve as validation targets for RQ2.

\textbf{Inter-rater disagreement.} With 3 annotators per scoring decision, each falls into one of two agreement categories: unanimous (3/3 agree), indicating an unambiguous decision, or split (2/1), indicating genuine ambiguity. When trained experts disagree on a binary judgment, it signals that the conversation genuinely straddles the boundary of the rubric criterion.

\begin{table}[H]
\centering
\textbf{Table 3.} Agreement distribution.

\smallskip
\begin{tabular}{ll}
\toprule
Metric & Value \\
\hline
Scoring decisions & 2,100 \\
Unanimous (3/3) & 1,887 (90\%) \\
Split (2/1) & 213 (10\%) \\
\bottomrule
\end{tabular}
\end{table}

\textbf{Annotator response time.} Response time offers a continuous measure of scoring difficulty: the cognitive effort required to judge a conversation should correlate with its ambiguity, as annotators spend more time deliberating on borderline cases. The raw response-time data has a heavy right tail (median 4.1 s, max \textgreater  24 h), likely reflecting breaks and idle or abandoned sessions; and annotators differ in pace (individual medians range from 2.6 s to 6.5 s). We process response times in three steps:

\begin{enumerate}
\item \textbf{Per-annotator percentile capping at p95} to remove likely breaks on a per-rater basis.
\item \textbf{Z-scoring within annotator} to control for individual speed differences.
\item \textbf{Median across the 3 annotators} as the per-decision measure, chosen over the mean to reduce sensitivity to any remaining outliers after capping.
\end{enumerate}

\textbf{Proxy correlation.} Before interpreting the two proxies independently, we check whether they measure the same underlying construct.

The two proxies are moderately correlated (\textit{r} = 0.323, \textit{p} \textless  .001): decisions where annotators disagree also take longer, with split decisions nearly a full standard deviation above unanimous ones (Cohen's \textit{d} = 0.93, \textit{p} \textless  .001; \citet{cohen1988statistical}). The proxies are correlated but not redundant --- disagreement captures outcome-level ambiguity while response time reflects deliberation effort even within unanimous decisions.

\subsection{Language models}

We test six LMs across three families, split into two size tiers. Small LMs return both a score and a confidence value in a single call. Large LMs return only a score, establishing the accuracy ceiling that the cascade aims to approach.

\textbf{Small LMs} (lower cost, return score + confidence):

\begin{itemize}
\item \texttt{gpt-5.4-nano} (\textbf{GPT Nano})
\item \texttt{claude-haiku-4.5} (\textbf{Claude Haiku})
\item \texttt{gemini-3.1-flash-lite-preview} (\textbf{Gemini Lite})
\end{itemize}

\textbf{Large LMs} (higher cost, return score only):

\begin{itemize}
\item \texttt{gpt-5.4} (\textbf{GPT})
\item \texttt{claude-opus-4.6} (\textbf{Claude Opus})
\item \texttt{gemini-3.1-pro-preview} (\textbf{Gemini Pro})
\end{itemize}

We use the bold short names throughout.

We estimate per-decision cost from each provider's published per-token pricing and the observed token counts in our scoring runs. We report per-model costs in Results.

\subsection{Scoring prompt}

Each LM receives a structured prompt containing the math problem, the full student-AI conversation, and the rubric criterion, and returns a JSON object with a binary scoring judgment and justification. Appendix~B reproduces the full prompt.

\subsection{Confidence elicitation}

Each small LM produces both a score and a confidence value in a single inference call. We append a \texttt{3\_Confidence} field to the existing scoring prompt's structured JSON output schema (the numeric prefix enforces generation order, ensuring the LM produces its scoring judgment and justification before stating confidence). The field instruction reads:

\begin{quote}
``What is the probability that your scoring judgment is correct? 0 = no confidence, 100 = certain.''
\end{quote}

Placing confidence after the scoring judgment ensures the LM commits to its decision before assessing uncertainty. Framing confidence as a probability follows \citet{yang2024verbalized}, who found this wording produced the best calibration for smaller LMs. We use task framing (describing the scoring task and rubric context) rather than human-persona prompts, which \citet{xu2025mirror} show can distort verbalized confidence.

\subsection{Inference settings}

All scoring calls use temperature T = 1 for consistency across models and tiers. GPT and Claude APIs require T = 1 when extended thinking is enabled, and Google recommends T = 1 for all Gemini calls. Extended thinking is disabled on all small LMs to minimize latency and cost, since the small LM's primary role is fast, cheap classification with a confidence signal. Large LMs use extended thinking at the ``high'' effort level to maximize scoring accuracy.

\textbf{Latency.} We record API response time (wall-clock latency) for every call. Cascade latency is sequential: the small LM always runs first, and escalated decisions incur small-plus-large LM latency. All scoring runs were completed within a single day; latency figures reflect API conditions during that window and may vary under different load.

\subsection{Cascade simulation design}

For each of the three family-matched pairs, both the small and large LM score all 2,100 decisions independently. We collect all scores upfront rather than making conditional API calls, which allows us to simulate cascade behavior at any threshold without adaptive effects (e.g., the large LM never sees the small LM's output).

Verbalized confidence values (0-100) are normalized to a [0, 1] scale. For a given threshold $\tau$, decisions where the small LM's confidence falls below $\tau$ are escalated: the small LM's score is replaced with the large LM's score, and all other decisions keep the small LM's score. We sweep every threshold from 0.01 to 0.99, each producing a different accuracy-cost combination. We then select an operating point in three steps:

\begin{enumerate}
\item Discard Pareto-dominated thresholds --- those where another option is both cheaper \textit{and} more accurate.
\item Among the remaining points, choose the cheapest threshold that keeps kappa within 0.02 of the large LM alone --- one-tenth of a Landis \& Koch interpretive band \citep{landis1977measurement}.
\item If no threshold meets the 0.02 criterion --- as happens when the small LM's confidence distribution is too narrow --- choose the Pareto-optimal point with the highest kappa. (This fallback applies when no threshold can approach large-LM accuracy, as occurs with Gemini Lite in our results.)
\end{enumerate}

Because this selection operates on the relative ranking of confidence scores rather than their absolute values, the procedure requires no assumptions about absolute calibration levels.

\subsection{Evaluation metrics}

\textbf{Discrimination (RQ1).} A small LM's prediction is accurate when it matches the human majority label. The primary metric is \textbf{AUROC} (area under the receiver operating characteristic curve): the probability that a randomly chosen accurate decision has higher confidence than a randomly chosen inaccurate one (0.5 = chance, 1.0 = perfect). Following \citet{hosmer2013assessing}, we interpret 0.7-0.8 as ``acceptable,'' 0.8-0.9 as ``excellent,'' and above 0.9 as ``outstanding.'' We report \textbf{ECE} (expected calibration error; \citet{guo2017calibration}) in the Appendix for comparison with prior work.

\textbf{Agreement (RQ1, RQ3).} We measure agreement between LM scores and human majority labels using Cohen's kappa \citep{cohen1960coefficient}, which corrects for chance agreement. We use kappa rather than raw accuracy because the class imbalance in our data would inflate accuracy.

\textbf{Cascade performance (RQ3).} For each cascade configuration, we report kappa, estimated cost per decision, and mean latency. We select the cheapest threshold that keeps kappa within 0.02 of the large LM alone.

\section{Results}

\subsection{Scoring accuracy}

We first establish each LM's baseline scoring accuracy, since the quality gap between small and large LMs determines whether a cascade can add value.

\begin{table}[H]
\centering
\textbf{Table 4.} Scoring accuracy by language model.

\smallskip
\begin{tabular}{lllll}
\toprule
LM & Tier & Accuracy & Cohen's kappa & 95\% CI \\
\hline
GPT Nano & Small & 88.2\% & 0.677 & [0.639, 0.714] \\
Claude Haiku & Small & 91.7\% & 0.783 & [0.751, 0.813] \\
Gemini Lite & Small & 89.5\% & 0.740 & [0.708, 0.772] \\
GPT & Large & 91.0\% & 0.770 & [0.739, 0.800] \\
Claude Opus & Large & 92.9\% & 0.819 & [0.791, 0.846] \\
Gemini Pro & Large & 92.5\% & 0.810 & [0.781, 0.838] \\
\bottomrule
\end{tabular}
\end{table}

All six LMs score well above chance. Among small LMs, Claude Haiku leads (91.7\% accuracy, kappa 0.783), followed by Gemini Lite (89.5\%, 0.740) and GPT Nano (88.2\%, 0.677). Among large LMs, Claude Opus and Gemini Pro score similarly (kappa 0.819 and 0.810). GPT is numerically lower (0.770), though its CI overlaps with both Claude Opus and Gemini Pro.

\subsection{RQ1: Confidence discrimination}

How well does verbalized confidence discriminate between accurate and inaccurate scores across the three small LMs?

\begin{table}[H]
\centering
\textbf{Table 5.} Confidence discrimination by small LM.

\smallskip
\begin{tabular}{llll}
\toprule
Small LM & AUROC & Distinct values & Variance \\
\hline
Claude Haiku & 0.857 & 13 & 0.0043 \\
Gemini Lite & 0.678 & 3 & 0.0005 \\
GPT Nano & 0.757 & 36 & 0.0057 \\
\bottomrule
\end{tabular}
\end{table}

\includegraphics[width=0.7\linewidth]{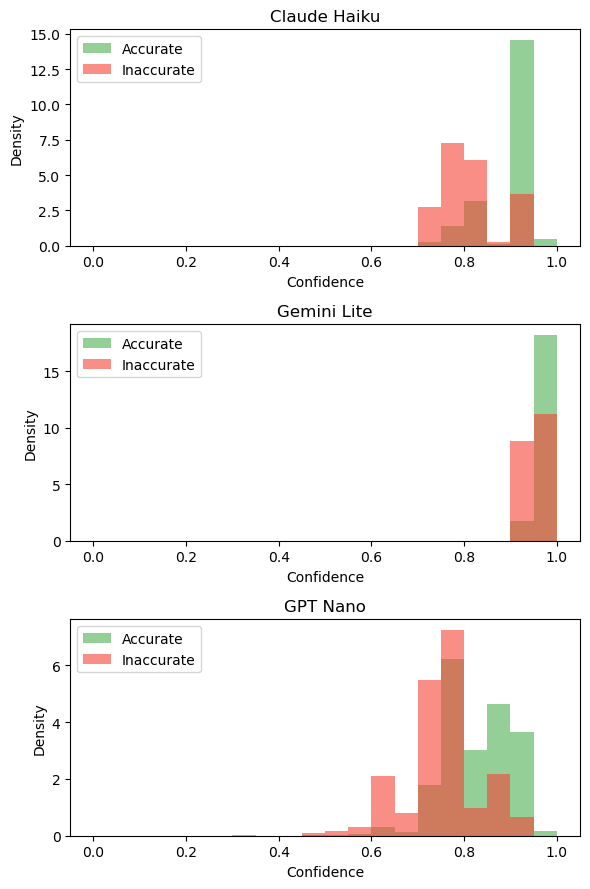}

\textbf{Figure 2.} Distribution of verbalized confidence for accurate (green) vs. inaccurate (red) predictions, by small LM. Claude Haiku shows clear separation between the two distributions; Gemini Lite clusters near 1.0 regardless of accuracy.

Claude Haiku produces the strongest confidence signal (AUROC 0.857, ``excellent'' discrimination per the Hosmer et al. scale), with clear separation visible in the histogram above. GPT Nano shows ``acceptable'' discrimination (AUROC 0.757) with the most granular scale (36 distinct values), though systematic underconfidence could trigger unnecessary escalation. Gemini Lite's confidence is severely limited for routing (AUROC 0.678): only 3 distinct values and near-zero variance (0.0005) mean the LM reports nearly the same confidence on every decision, matching the flat difficulty-sensitivity pattern that \citet{xu2025mirror} document.

For the cascade simulations in RQ3, Claude Haiku-based routing has the strongest discriminative foundation; Gemini Lite's near-degenerate confidence will limit what any threshold can achieve. Appendix~A makes this contrast visually stark and provides calibration metrics and stratified results by agreement status.

\subsection{RQ2: AI confidence and human uncertainty}

We validate confidence against the two human uncertainty proxies: inter-rater disagreement and annotator response time.

\subsubsection{Inter-rater disagreement}

If confidence tracks genuine scoring difficulty, LMs should report lower confidence on split decisions than on unanimous ones. A larger gap indicates greater sensitivity to scoring ambiguity.

\begin{table}[H]
\centering
\textbf{Table 6.} Mean confidence by human agreement status.

\smallskip
\begin{tabular}{lllll}
\toprule
Small LM & Unanimous & Split & Cohen's \textit{d} & \textit{p} \\
\hline
Claude Haiku & 0.909 & 0.833 & 1.16 & \textless  .001 \\
Gemini Lite & 0.994 & 0.980 & 0.53 & \textless  .001 \\
GPT Nano & 0.824 & 0.769 & 0.69 & \textless  .001 \\
\bottomrule
\end{tabular}
\end{table}

\includegraphics[width=0.7\linewidth]{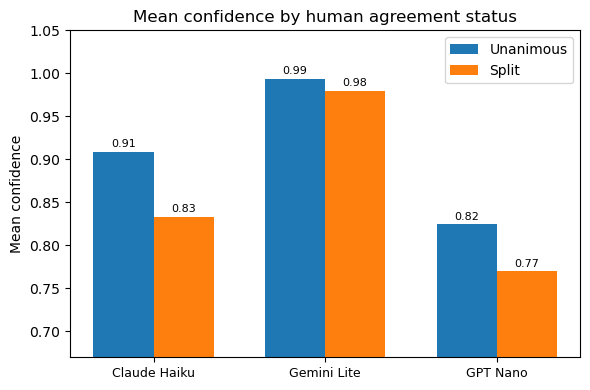}

\textbf{Figure 3.} Mean verbalized confidence for unanimous vs. split scoring decisions, by small LM. A larger gap indicates greater sensitivity to human scoring difficulty.

All three small LMs report lower confidence on split decisions (all \textit{p} \textless  .001), but the effect sizes vary widely. Claude Haiku shows the largest gap (\textit{d} = 1.16): its confidence drops meaningfully when humans disagree, making it the most sensitive to scoring ambiguity. GPT Nano shows a moderate gap (\textit{d} = 0.69). Gemini Lite's near-degenerate confidence (see RQ1) leaves almost no room to distinguish unanimous from split cases (\textit{d} = 0.53).

\subsubsection{Annotator response time}

We also test whether confidence tracks a second proxy for difficulty: annotator response time.

\begin{table}[H]
\centering
\textbf{Table 7.} Confidence vs. annotator response time.

\smallskip
\begin{tabular}{lll}
\toprule
Small LM & Spearman's rho & \textit{p} \\
\hline
Claude Haiku & -0.427 & \textless  .001 \\
Gemini Lite & -0.175 & \textless  .001 \\
GPT Nano & -0.267 & \textless  .001 \\
\bottomrule
\end{tabular}
\end{table}

All three small LMs show the expected negative correlation: higher confidence corresponds to faster human scoring times. Claude Haiku shows the strongest association (rho = -0.43), consistent with its superior sensitivity to scoring difficulty. Gemini Lite's correlation is weakest (rho = -0.18), consistent with the limited variance noted in RQ1.

\subsection{Per-model cost}

Before evaluating cascade savings, we report the estimated cost per scoring decision for each model, based on observed token counts and published per-token pricing.

\begin{table}[H]
\centering
\textbf{Table 8.} Estimated cost per 1,000 scoring decisions by model.

\smallskip
\begin{tabular}{lll}
\toprule
LM & Tier & Cost per 1k decisions \\
\hline
GPT Nano & Small & \$0.30 \\
Claude Haiku & Small & \$1.64 \\
Gemini Lite & Small & \$0.39 \\
GPT & Large & \$6.88 \\
Claude Opus & Large & \$9.49 \\
Gemini Pro & Large & \$5.59 \\
\bottomrule
\end{tabular}
\end{table}

\subsection{RQ3: Cascade performance}

The results above establish that verbalized confidence carries signal about decision difficulty --- but discrimination and human-difficulty alignment are necessary, not sufficient. The cascade must also deliver cost and latency savings in practice.

\subsubsection{Threshold optimization}

We sweep confidence thresholds from 0.01 to 0.99. At each threshold, decisions where the small LM's confidence falls below that value are escalated to the large LM. Higher thresholds escalate more decisions, improving accuracy but increasing cost. We select the cheapest Pareto-optimal threshold that keeps kappa within 0.02 of the large LM alone (we verify this introduces negligible optimism via five-fold cross-validation; see Appendix~C). We report bootstrap 95\% confidence intervals for all kappa estimates.

\includegraphics[width=0.7\linewidth]{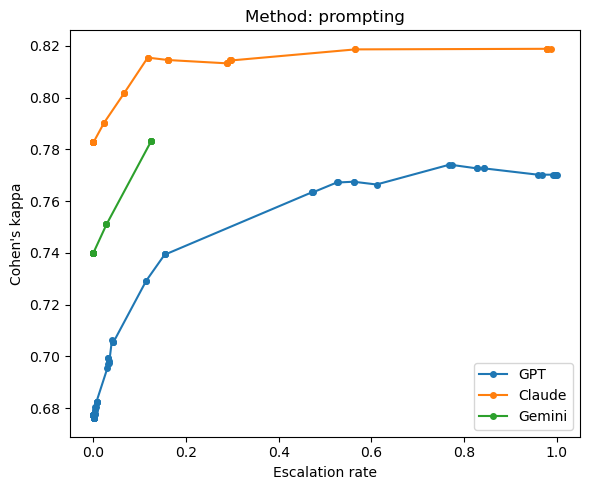}

\textbf{Figure 4.} Kappa-escalation rate tradeoff for each model family. Each point represents a confidence threshold; higher escalation rates improve agreement at greater cost. Note: the y-axis is compressed (range 0.68--0.82) to show the tradeoff curve clearly.

\begin{table}[H]
\centering
\textbf{Table 9.} Optimal cascade thresholds and performance.

\smallskip
\begin{tabular}{lllll}
\toprule
Family & $\tau$ & Kappa & 95\% CI & Esc. rate \\
\hline
Claude & 0.77 & 0.802 & [0.77,~0.83] & 7\% \\
GPT & 0.80 & 0.763 & [0.73,~0.79] & 47\% \\
Gemini & 0.99 & 0.783 & [0.75,~0.81] & 13\% \\
\bottomrule
\end{tabular}
\end{table}

At these thresholds, the Claude cascade reduces cost by 76\% and median latency by 61\% versus always-large scoring; GPT reduces cost by 49\% and latency by 64\%; Gemini reduces cost by 81\% and latency by 82\%.

To quantify uncertainty in the cascade's kappa loss, we bootstrap the paired kappa difference (cascade minus large-alone) on the same scoring decisions.

\begin{table}[H]
\centering
\textbf{Table 10.} Cascade vs. large-alone kappa difference (paired bootstrap, 10k resamples).

\smallskip
\begin{tabular}{lll}
\toprule
Family & Kappa difference & 95\% CI \\
\hline
Claude & -0.017 & [-0.039, +0.005] \\
GPT & -0.007 & [-0.020, +0.005] \\
Gemini & -0.026 & [-0.042, -0.010] \\
\bottomrule
\end{tabular}
\end{table}

The GPT and Claude difference CIs include zero (-0.007 [-0.020, +0.005] and -0.017 [-0.039, +0.005], respectively), meaning neither cascade's kappa is statistically distinguishable from the large LM alone --- though the Claude CI is wide enough to accommodate a loss of up to 0.039, so practitioners should weigh this uncertainty against their tolerance for accuracy degradation. Gemini's CI falls entirely below zero (-0.026 [-0.042, -0.010]), reflecting Gemini Lite's poor calibration --- the cascade cannot close the quality gap when the routing signal is uninformative.

We select the threshold on the evaluation data; deployment would require selecting $\tau$ on a held-out calibration set. Five-fold cross-validation confirms negligible optimism from this design (see Appendix~C).

\subsubsection{Cost-accuracy tradeoff}

\includegraphics[width=0.7\linewidth]{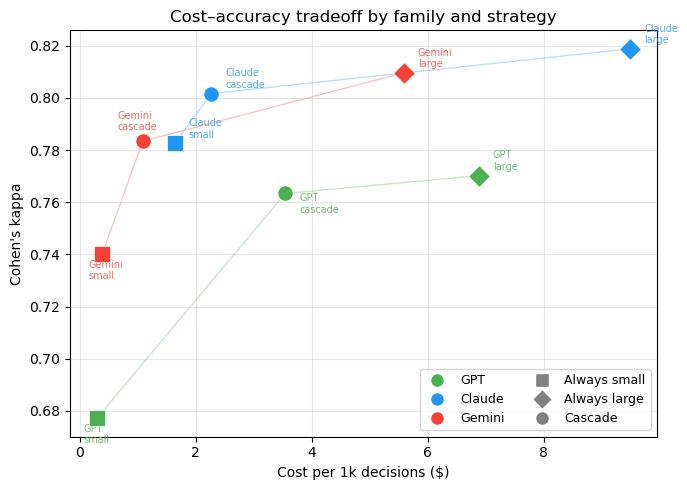}

\textbf{Figure 5.} Cost-accuracy tradeoff for each model family. Cascade systems (circles) approach always-large accuracy (diamonds) at near-always-small cost (squares). Lines connect strategies within each family.

The cascade positions each family between its always-small and always-large baselines. Claude comes closest to always-large accuracy at near-always-small cost (7\% escalation rate). GPT sits mid-range due to its higher escalation rate. Gemini achieves the largest cost reduction (81\%) but cannot close the accuracy gap, consistent with Gemini Lite's poor calibration.

\subsubsection{Escalation lift}

For routing to improve outcomes, two conditions must hold: (1) low-confidence cases must genuinely be harder --- the small LM must agree less with human labels on escalated cases than on cases it keeps --- and (2) the large LM must do better on those hard cases than the small LM did. We test both conditions by splitting each family's decisions at the Pareto-optimal threshold and comparing small vs. large LM kappa on the escalated subset.

\begin{table}[H]
\centering
\textbf{Table 11.} Confidence separation: small-LM kappa on kept vs. escalated decisions.

\smallskip
\begin{tabular}{llll}
\toprule
Family & Kappa (kept) & Kappa (escalated) & Separation \\
\hline
GPT & 0.869 & 0.502 & +0.367 \\
Claude & 0.831 & 0.211 & +0.620 \\
Gemini & 0.828 & 0.250 & +0.578 \\
\bottomrule
\end{tabular}
\end{table}

\begin{table}[H]
\centering
\textbf{Table 12.} Large-LM lift on escalated decisions.

\smallskip
\begin{tabular}{lllll}
\toprule
Family & N escalated & Small kappa (esc) & Large kappa (esc) & Lift \\
\hline
GPT & 990 & 0.502 & 0.669 & +0.167 \\
Claude & 139 & 0.211 & 0.471 & +0.261 \\
Gemini & 263 & 0.250 & 0.513 & +0.263 \\
\bottomrule
\end{tabular}
\end{table}

All three families show confidence correctly identifying harder cases, with small-LM kappa dropping by 0.37--0.62 on escalated decisions --- from ``almost perfect'' agreement to ``moderate'' or ``fair'' agreement. The large LM improves on those cases in all families, with kappa lift ranging from +0.17 (GPT) to +0.26 (Gemini). GPT's high escalation rate (47\%) dilutes this advantage by including moderately easy cases; Claude and Gemini escalate more selectively and show larger lift on the escalated subset, because their smaller escalated pools concentrate a higher share of genuinely hard cases.

\subsubsection{Latency profiles}

For cascade routing to reduce mean latency, the speed gain on non-escalated decisions must outweigh the cost of escalated ones running both LMs sequentially. We compare latency profiles across the three families.

\begin{table}[H]
\centering
\textbf{Table 13.} Latency profiles by scoring strategy.

\smallskip
\begin{tabular}{llll}
\toprule
Family & Strategy & Median (ms) & p95 (ms) \\
\hline
Claude & Always large & 5,300 & 14,806 \\
Claude & Always small & 2,049 & 2,888 \\
Claude & Cascade & 2,066 & 8,228 \\
GPT & Always large & 3,370 & 11,236 \\
GPT & Always small & 700 & 1,229 \\
GPT & Cascade & 1,199 & 10,519 \\
Gemini & Always large & 9,118 & 36,196 \\
Gemini & Always small & 1,512 & 2,097 \\
Gemini & Cascade & 1,617 & 14,355 \\
\bottomrule
\end{tabular}
\end{table}

\includegraphics[width=0.7\linewidth]{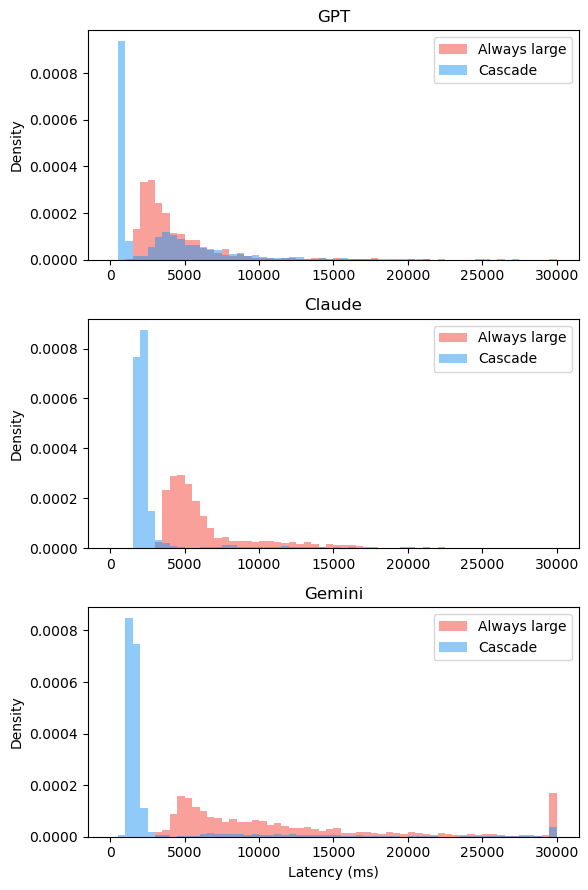}

\textbf{Figure 6.} Latency distributions for always-large vs. confidence cascade scoring, by model family. The cascade produces a bimodal distribution: a fast mode from small-LM-only decisions and a slow tail from escalated decisions.

All three cascades reduce median latency compared to always-large scoring, ranging from 61\% (Claude) to 82\% (Gemini). The spread reflects the speed gap between tiers: Gemini Pro is exceptionally slow relative to Gemini Lite, while Claude Haiku and Claude Opus are closer in speed. GPT achieves a 64\% reduction despite escalating 47\% of decisions, because GPT Nano is fast enough that even sequential calls finish well under the always-large median. Cascade p95 latency is lower than always-large p95 in all families, though the bimodal mix of fast kept decisions and slow escalated ones produces a heavy tail.

\section{Discussion}

\subsection{Confidence signal quality as the bottleneck}

Prior work has shown cost savings from LM cascades on general NLP tasks \citep{chen2023frugalgpt, gupta2024cascades, zellinger2024hcma}. The central finding here is that the quality of the confidence signal --- not model accuracy alone --- is the bottleneck determining whether a cascade works well. A cascade routes decisions by comparing confidence to a threshold, so the signal must discriminate between accurate and inaccurate decisions and vary enough for a threshold to act on. The three small LMs differ sharply on both counts, and cascade kappa loss tracks this directly: Claude Haiku (AUROC 0.857, 13 distinct values) yields a cascade with no detectable kappa loss, while Gemini Lite (AUROC 0.678, 3 distinct values near 1.0) yields the largest loss, with its CI entirely below zero. Our data cannot disentangle the two properties; the LMs with poor discrimination also lack spread --- but both are necessary for a threshold to work.

Calibration (whether stated probabilities match observed accuracy) matters for a separate reason. Well-calibrated values carry a stable meaning across settings, so a threshold tuned in one context is more likely to transfer. With poor calibration, practitioners will likely need labeled data to re-tune thresholds for each new deployment. (Appendix~A reports absolute calibration metrics.)

Claude Haiku's AUROC of 0.857 exceeds the best previously reported verbalized-confidence results we are aware of: \citet{tao2025revisiting} find AUROC between 0.7 and 0.8 for GPT-4.1 and Grok 3 on reasoning benchmarks. This comparison spans different LMs, tasks, and domains, so the gap could reflect task difficulty or model-specific factors rather than a general advantage for educational scoring.

\subsection{What confidence tracks}

The RQ2 results show that confidence correlates with human-judged difficulty, but they do not establish that confidence \textit{captures} ambiguity in any deep sense. The correlation could arise from surface features shared between LM uncertainty and human difficulty: conversation length, vocabulary complexity, or rubric-criterion overlap, rather than from the LM modeling the same ambiguity that humans experience. For example, longer conversations may both lower LM confidence (more tokens increase the chance of conflicting signals) and take annotators longer to score, creating a spurious correlation even if the LM is not tracking ambiguity per se. What we can say is that, whatever its source, the confidence signal is informative enough to separate easy from hard cases for routing purposes, and that this separation aligns with independent human difficulty proxies.

\subsection{When cascades help, and when they don't}

The agreement lift from escalation depends on two factors that interact: how selectively the small LM escalates, and how large the quality gap is between tiers. Both must be present. Gemini Lite has the largest accuracy gap to close, but its near-degenerate confidence signal means the cascade cannot identify which cases to escalate; the cost--accuracy tradeoff plot shows it saving cost but never closing the accuracy gap. Conversely, when the small LM is already highly accurate (Claude Haiku), a strong confidence signal lets the cascade match large-LM accuracy at near-small-LM cost, even though the room for accuracy improvement is narrow.

GPT illustrates a third pattern: what happens when confidence discriminates but is systematically underconfident. GPT Nano's 47\% escalation rate means nearly half of all decisions reach the large LM. On accuracy, this is benign --- the cascade's kappa CI still includes zero. On efficiency, it is costly: escalating half the decisions gives up most of the latency and cost savings that make cascades attractive in the first place.

More broadly, a strong confidence signal gives practitioners a smooth Pareto frontier to dial between cost and accuracy; a weak one collapses the frontier to a single extreme point with no room to tune.

\subsection{Limitations}

This study has six scope constraints:

\begin{itemize}
\item \textbf{Narrow item coverage.} Seven criteria across four math items, all from conversation-based math assessments. Calibration patterns may differ for other item types, subject areas, or assessment formats.
\item \textbf{High-agreement criteria only.} All criteria have substantial human agreement (kappa at or above 0.70). Confidence-based routing may behave differently on criteria with lower human agreement, where ground truth is noisier and routing decisions are harder to validate.
\item \textbf{Binary rubrics only.} Multi-level rubrics may interact differently with confidence elicitation.
\item \textbf{Single confidence prompt.} Verbalized confidence is sensitive to prompting \citep{yang2024verbalized}. Our calibration results may not generalize to alternative phrasings, different placement of the confidence field within the response schema, or different framing (e.g., asking for uncertainty rather than confidence). Prompt robustness testing is a prerequisite for production deployment.
\item \textbf{No logprob-based confidence.} Availability is inconsistent across providers: Claude does not expose logprobs, and GPT provides them only with reasoning disabled. Future work could compare logprobs against prompting where provider support permits.
\item \textbf{Threshold selection on evaluation data.} We select the Pareto-optimal threshold on the same data used for evaluation. Five-fold cross-validation shows negligible optimism from this design (see Appendix~C), but deployment would require selecting thresholds on a held-out calibration set.
\end{itemize}

LM capabilities evolve rapidly, and results apply to the specific LM versions tested. Testing across three families provides some evidence of generalizability.

\subsection{Implications}

Beyond cost savings, cascade routing may make synchronous scoring more viable. In conversation-based assessments where each student turn requires a scoring judgment before the system can respond, or in adaptive tests where item selection depends on the current score, the scorer sits in the critical path of the student experience. The cascade reduces median latency by 61--82\% across the three families tested, with most decisions returning at small-LM speed and only the uncertain minority paying extra time for escalation. Without cascade routing, synchronous scoring with large LMs introduces per-turn delays that degrade the assessment experience. The cost savings are substantial, but the latency reduction may be more consequential: it is the difference between an architecture that works in real time and one that does not.

Verbalized confidence also has practical advantages over alternative confidence methods. It works with any LM API, with no dependency on logprob access (which varies across providers and model types), making it viable for organizations that switch or multi-source providers. It adds no extra API calls: the confidence value arrives in the same response as the score, preserving the cost reduction that motivates the cascade in the first place. And it requires no fine-tuning, probe training, or internal model access; the signal is part of the prompt--response contract, making it auditable and straightforward to implement.

Confidence signals may also have diagnostic value beyond routing. Items or criteria that consistently elicit low LM confidence could flag rubric language that needs revision --- an open question for future work.

\subsection{Conclusion}

The quality of the small LM's confidence signal --- not model accuracy alone --- determines whether a cascade delivers on its promise. When the signal discriminates well and the accuracy gap between tiers is large enough to justify escalation, verbalized confidence enables near-large-LM accuracy at a fraction of the cost and latency: the Claude cascade matched large-LM kappa at 76\% lower cost and 61\% lower latency. When either condition fails, as with Gemini Lite's near-degenerate confidence distribution, the cascade cannot identify which cases to escalate and the savings come at the expense of accuracy. Because verbalized confidence requires no logprob access, no fine-tuning, and no extra API calls, it is a practical routing signal for any provider --- but practitioners should verify discrimination on their task before relying on it.

\section{Appendices}

\subsection{Appendix A: Calibration}\label{appendix-a}

The main body evaluates confidence via AUROC (discrimination). Here we report absolute calibration --- how closely stated confidence tracks actual accuracy. This matters for applications that use raw confidence values, such as reporting uncertainty to teachers or setting thresholds without labeled validation data.

\begin{table}[H]
\centering
\textbf{Table 14.} Absolute calibration metrics by small LM.

\smallskip
\begin{tabular}{lll}
\toprule
Small LM & ECE & NCE \\
\hline
Claude Haiku & 0.042 & +0.016 \\
Gemini Lite & 0.097 & -0.097 \\
GPT Nano & 0.073 & +0.063 \\
\bottomrule
\end{tabular}
\end{table}

Claude Haiku's ECE of 0.042 indicates strong absolute calibration --- stated confidence closely tracks actual accuracy. Its positive NCE (+0.016) confirms slight underconfidence. GPT Nano has moderate calibration with stronger underconfidence (NCE +0.063). Gemini Lite's ECE reflects near-ceiling confidence on all decisions rather than meaningful calibration.

The reliability diagrams \citep{degroot1983comparison} below show the same data visually. Each bar represents a confidence bin: the x-axis is the LM's stated confidence and the y-axis is actual accuracy in that bin. A perfectly calibrated LM falls on the dashed diagonal (80\% confidence = 80\% accuracy). Bars above the line indicate underconfidence; bars below indicate overconfidence.

\includegraphics[width=0.7\linewidth]{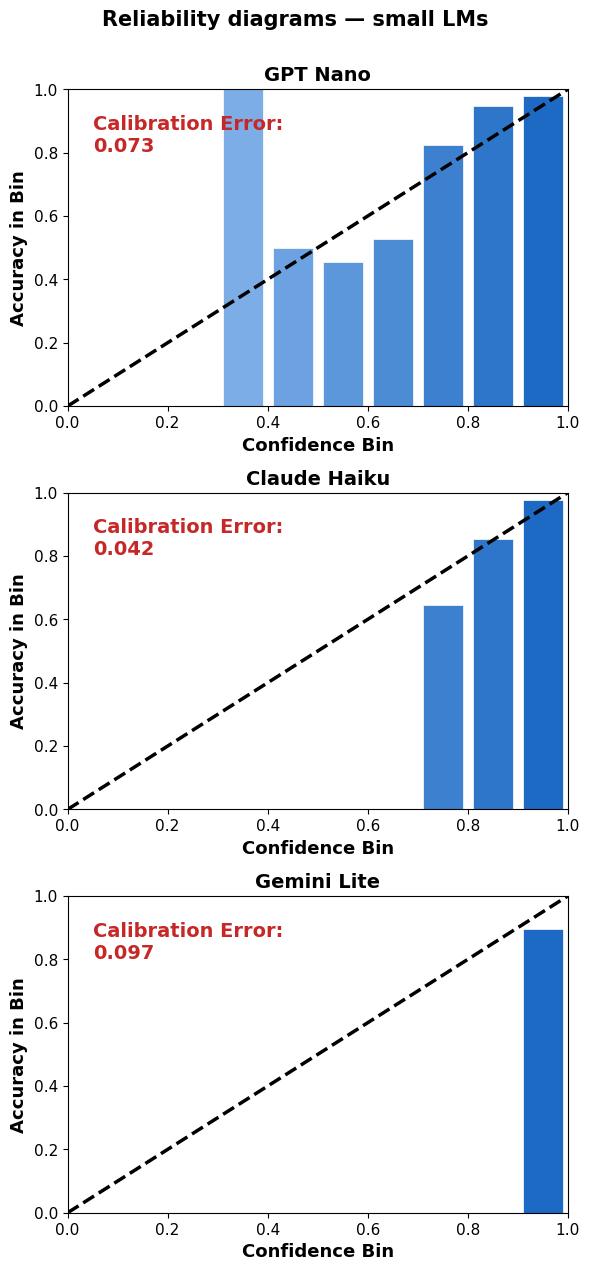}

\textbf{Figure 7.} Reliability diagrams for each small LM. Bars show actual accuracy per confidence bin; the dashed diagonal represents perfect calibration. Bars above the line indicate underconfidence.

\subsubsection{By ground-truth certainty}

Ground truth is less certain for split decisions (where annotators disagreed 2/1) than for unanimous ones. Calibration metrics computed against majority vote therefore embed measurement noise in the split subset. If verbalized confidence is sensitive to genuine ambiguity, we should see degraded calibration on split cases --- the LM may be appropriately uncertain rather than miscalibrated.

\begin{table}[H]
\centering
\textbf{Table 15.} Calibration metrics stratified by ground-truth certainty.

\smallskip
\begin{tabular}{lllll}
\toprule
Small LM & Subset & N & AUROC & ECE \\
\hline
Claude Haiku & unanimous & 1,887 & 0.876 & 0.051 \\
Claude Haiku & split & 213 & 0.620 & 0.209 \\
 &  &  &  &  \\
Gemini Lite & unanimous & 1,887 & 0.720 & 0.066 \\
Gemini Lite & split & 213 & 0.512 & 0.374 \\
 &  &  &  &  \\
GPT Nano & unanimous & 1,887 & 0.778 & 0.094 \\
GPT Nano & split & 213 & 0.595 & 0.180 \\
\bottomrule
\end{tabular}
\end{table}

Calibration degrades sharply on split decisions across all three small LMs. Claude Haiku's AUROC drops from 0.876 (unanimous) to 0.620 (split), and its ECE rises from 0.051 to 0.209. Gemini Lite's split AUROC (0.512) is near chance. GPT Nano falls between the two. On decisions where even trained annotators disagree, the majority label is an imperfect proxy for correctness, and the LM's uncertainty is at least partly appropriate. The unanimous subset --- 90\% of decisions --- shows that verbalized confidence works well when ground truth is clear.

\subsection{Appendix B: Scoring prompt}\label{appendix-b}

The prompt below is the verbalized-confidence variant used for small LMs. It adds the \texttt{3\_Confidence} field to the base scoring prompt. Large LMs receive the same prompt without the confidence field. We fill template variables (\texttt{\{problem\}}, \texttt{\{student\_answer\}}, \texttt{\{criterion\}}, \texttt{\{conversation\}}) per-decision at inference time.

\begin{verbatim}
<Role>
You are an evaluator reviewing a conversation
between an AI assessment proctor and a student.
Determine whether the student satisfied the
evaluation criterion based on their responses.
Do NOT continue the conversation. Do NOT
roleplay as the proctor or the student.
</Role>

<Problem>{problem}</Problem>
<StudentAnswer>{student_answer}</StudentAnswer>
<Criterion>{criterion}</Criterion>
<Conversation>
{conversation}
</Conversation>

<Requirements>
 - Do NOT mark the criterion as satisfied unless
  you are confident that the student has
  demonstrated understanding of the criterion.
 - Students do not need to use the EXACT concept
  terms in the criterion, but can use synonymous
  language.
 - Evaluate based on the full context: the
  problem, the student's answer, and their
  conversation.
 - Do not assume the student understands the
  criterion based on a correct answer to the
  Problem.
 - Don't mark the criterion as satisfied if the
  student is describing a step in solving the
  problem (unless that is necessary to
  demonstrate understanding of the criterion).
</Requirements>

Respond with a JSON object matching this schema:

{
  "1_Reasoning": "Brief reasoning about whether
    the criterion is satisfied (25 words max).",
  "2_IsSatisfied": "<true or false>",
  "3_Confidence": "<integer 0 -100>"
}

 - 1_Reasoning (string, required): Brief
  reasoning about whether the student has
  satisfied the criterion. Keep to 25 words
  at most.
 - 2_IsSatisfied (boolean, required): true if
  the criterion has been satisfied, false
  otherwise.
 - 3_Confidence (integer, required): What is the
  probability that your scoring judgment is
  correct? 0 = no confidence, 100 = certain.

Return ONLY the JSON object, no other text.
\end{verbatim}

\subsection{Appendix C: Cross-validated threshold selection}\label{appendix-c}

The main results select the cascade threshold on the full evaluation set. To quantify the optimism this introduces, we run five-fold stratified cross-validation. Each fold selects the threshold on 80\% of scoring decisions using the same Pareto-optimal selection rule, then evaluates cascade kappa on the held-out 20\%. We report the mean and standard deviation across folds.

\begin{table}[H]
\centering
\textbf{Table 16.} In-sample vs. cross-validated cascade performance.

\smallskip
\begin{tabular}{lllll}
\toprule
Family & In-sample kappa & CV kappa (SD) & Optimism & Thr. SD \\
\hline
GPT & 0.763 & 0.764 (0.048) & -0.000 & 0.00 \\
Claude & 0.802 & 0.804 (0.029) & -0.002 & 0.02 \\
Gemini & 0.783 & 0.784 (0.042) & -0.000 & 0.00 \\
\bottomrule
\end{tabular}
\end{table}

In-sample and cross-validated kappa agree closely across all three families. Optimism is negligible --- the largest difference is less than 0.005 kappa --- and threshold selection is stable across folds, confirming that in-sample threshold selection does not overfit in this setting.

%%%%%%%%%%%%%%   Bibliography   %%%%%%%%%%%%%%
\bibliography{main.bib}

@misc{openai2026pricing,
	author = {{OpenAI}},
	year = {2026},
	note = {Accessed 2026-03-22},
	title = {API {Pricing}},
	url = {https://developers.openai.com/api/docs/pricing},
	howpublished = {https://developers.openai.com/api/docs/pricing},
}

@inproceedings{xiong2024confidence,
	author = {Xiong, Miao and Hu, Zhiyuan and Lu, Xinyang and Li, Yifei and Fu, Jie and He, Junxian and Hooi, Bryan},
	booktitle = {International {Conference} on {Learning} {Representations} ({ICLR})},
	doi = {10.48550/arXiv.2306.13063},
	year = {2024},
	note = {arXiv:2306.13063},
	title = {Can {LLMs} {Express} {Their} {Uncertainty}? {An} {Empirical} {Evaluation} of {Confidence} {Elicitation} in {LLMs}},
}

@article{tao2025revisiting,
	author = {Tao, Linwei and Yeh, Yi-Fan and Dong, Minjing and Huang, Tao and Torr, Philip and Xu, Chang},
	doi = {10.48550/arXiv.2505.23854},
	year = {2025},
	note = {arXiv:2505.23854},
	title = {Revisiting {Uncertainty} {Estimation} and {Calibration} of {Large} {Language} {Models}},
}

@article{tao2025human,
	author = {Tao, Linwei and Yeh, Yi-Fan and Kai, Bo and Dong, Minjing and Huang, Tao and Lamb, Tom A. and Yu, Jialin and Torr, Philip H. S. and Xu, Chang},
	doi = {10.48550/arXiv.2509.24202},
	year = {2025},
	note = {arXiv:2509.24202},
	title = {Can {Large} {Language} {Models} {Express} {Uncertainty} {Like} {Human}?},
}

@inproceedings{frohn2025scoring,
	address = {Palermo, Italy},
	author = {Frohn, Scott and Burleigh, Tyler and Chen, Jing},
	series = {Lecture {Notes} in {Artificial} {Intelligence}},
	booktitle = {Artificial {Intelligence} in {Education}},
	doi = {10.1007/978-3-031-98465-5_6},
	year = {2025},
	pages = {44--51},
	organization = {Springer},
	title = {Automated {Scoring} of {Short} {Answer} {Questions} with {Large} {Language} {Models}: Impacts of {Model}, {Item}, and {Rubric} {Design}},
	volume = {VI},
}

@inbook{burstein2013erater,
	author = {Burstein, Jill and Tetreault, Joel and Madnani, Nitin},
	booktitle = {Handbook of {Automated} {Essay} {Evaluation}: Current {Applications} and {New} {Directions}},
	doi = {10.4324/9780203122761},
	editor = {Shermis, Mark D. and Burstein, Jill C.},
	year = {2013},
	pages = {55--67},
	publisher = {Routledge},
	title = {The e-rater\textregistered{} {Automated} {Essay} {Scoring} {System}},
}

@article{ramineni2013guidelines,
	author = {Ramineni, Chaitanya and Williamson, David M.},
	journal = {Assessing Writing},
	doi = {10.1016/j.asw.2012.10.004},
	number = {1},
	year = {2013},
	pages = {25--39},
	title = {Automated {Essay} {Scoring}: Psychometric {Guidelines} and {Practices}},
	volume = {18},
}

@inbook{foltz2013iea,
	author = {Foltz, Peter W. and Streeter, Lynn A. and Lochbaum, Karen E. and Landauer, Thomas K.},
	booktitle = {Handbook of {Automated} {Essay} {Evaluation}: Current {Applications} and {New} {Directions}},
	doi = {10.4324/9780203122761},
	editor = {Shermis, Mark D. and Burstein, Jill C.},
	year = {2013},
	pages = {68--88},
	publisher = {Routledge},
	title = {Automated {Scoring} of {Essays} with the {Intelligent} {Essay} {Assessor}},
}

@inproceedings{funayama2020cse,
	author = {Funayama, Hiroaki and Sasaki, Shota and Matsubayashi, Yuichiroh and Mizumoto, Tomoya and Suzuki, Jun and Mita, Masato and Inui, Kentaro},
	booktitle = {Proceedings of the 58th {Annual} {Meeting} of the {Association} for {Computational} {Linguistics}: Student {Research} {Workshop}},
	doi = {10.18653/v1/2020.acl-srw.32},
	year = {2020},
	pages = {237--243},
	organization = {Association for Computational Linguistics},
	title = {Preventing {Critical} {Scoring} {Errors} in {Short} {Answer} {Scoring} with {Confidence} {Estimation}},
}

@inproceedings{funayama2022hitl,
	author = {Funayama, Hiroaki and Sato, Tasuku and Matsubayashi, Yuichiroh and Mizumoto, Tomoya and Suzuki, Jun and Inui, Kentaro},
	booktitle = {International {Conference} on {Artificial} {Intelligence} in {Education}},
	doi = {10.48550/arXiv.2206.08288},
	year = {2022},
	note = {arXiv:2206.08288},
	pages = {465--476},
	organization = {Springer},
	title = {Balancing {Cost} and {Quality}: An {Exploration} of {Human}-in-the-{Loop} {Frameworks} for {Automated} {Short} {Answer} {Scoring}},
}

@inproceedings{xiao2025dualprocess,
	author = {Xiao, Changrong and Ma, Wenxing and Song, Qingping and Xu, Sean Xin and Zhang, Kunpeng and Wang, Yufang and Fu, Qi},
	booktitle = {Proceedings of the 15th {International} {Learning} {Analytics} and {Knowledge} {Conference}},
	doi = {10.1145/3706468.3706507},
	year = {2025},
	pages = {293--305},
	title = {Human-{AI} {Collaborative} {Essay} {Scoring}: A {Dual}-{Process} {Framework} with {LLMs}},
}

@inproceedings{geng2024survey,
	author = {Geng, Jiahui and Cai, Fengyu and Wang, Yuxia and Koeppl, Heinz and Nakov, Preslav and Gurevych, Iryna},
	booktitle = {Proceedings of the 2024 {Conference} of the {North} {American} {Chapter} of the {Association} for {Computational} {Linguistics} ({NAACL})},
	doi = {10.18653/v1/2024.naacl-long.366},
	year = {2024},
	pages = {6577--6595},
	title = {A {Survey} of {Confidence} {Estimation} and {Calibration} in {Large} {Language} {Models}},
}

@article{chen2023frugalgpt,
	author = {Chen, Lingjiao and Zaharia, Matei and Zou, James},
	doi = {10.48550/arXiv.2305.05176},
	year = {2023},
	note = {arXiv:2305.05176},
	title = {FrugalGPT: How to {Use} {Large} {Language} {Models} {While} {Reducing} {Cost} and {Improving} {Performance}},
}

@inproceedings{chuang2025routing,
	author = {Chuang, Yu-Neng and Yu, Leisheng and Wang, Guanchu and Zhang, Lizhe and Liu, Zirui and Cai, Xuanting and Sui, Yang and Braverman, Vladimir and Hu, Xia},
	doi = {10.48550/arXiv.2502.04428},
	year = {2025},
	note = {arXiv:2502.04428},
	title = {Confident or {Seek} {Stronger}: Exploring {Uncertainty}-{Based} {On}-{Device} {LLM} {Routing}},
}

@inproceedings{burleigh2025prepilot,
	author = {Burleigh, Tyler and Chen, Jing and DiCerbo, Kristen},
	booktitle = {Proceedings of the {Artificial} {Intelligence} in {Measurement} and {Education} {Conference} ({AIME}-{Con}): Coordinated {Session} {Papers}},
	isbn = {979-8-218-84230-7},
	year = {2025},
	pages = {61--68},
	organization = {National Council on Measurement in Education (NCME)},
	title = {Pre-{Pilot} {Optimization} of {Conversation}-{Based} {Assessment} {Items} {Using} {Synthetic} {Response} {Data}},
	url = {https://aclanthology.org/2025.aimecon-sessions.7/},
}

@inproceedings{yildirim2023cba,
	author = {Yildirim-Erbasli, Seyma and Bulut, Okan},
	booktitle = {2023 {IEEE} {International} {Conference} on {Advanced} {Learning} {Technologies} ({ICALT})},
	doi = {10.1109/ICALT58122.2023.00103},
	year = {2023},
	pages = {331--335},
	title = {Innovating {Assessment} with {Conversational} {Agents}: A {Technology}-{Enhanced} {Approach} to {Formative} {Assessments}},
}

@article{fleiss1971measuring,
	author = {Fleiss, Joseph L.},
	journal = {Psychological Bulletin},
	doi = {10.1037/h0031619},
	number = {5},
	year = {1971},
	pages = {378--382},
	title = {Measuring {Nominal} {Scale} {Agreement} {Among} {Many} {Raters}},
	volume = {76},
}

@article{landis1977measurement,
	author = {Landis, J. Richard and Koch, Gary G.},
	journal = {Biometrics},
	doi = {10.2307/2529310},
	number = {1},
	year = {1977},
	pages = {159--174},
	title = {The {Measurement} of {Observer} {Agreement} for {Categorical} {Data}},
	volume = {33},
}

@book{cohen1988statistical,
	author = {Cohen, Jacob},
	edition = {2nd},
	isbn = {978-0-8058-0283-2},
	year = {1988},
	publisher = {Lawrence Erlbaum Associates},
	title = {Statistical {Power} {Analysis} for the {Behavioral} {Sciences}},
}

@article{yang2024verbalized,
	author = {Yang, Daniel and Tsai, Yao-Hung Hubert and Yamada, Makoto},
	doi = {10.48550/arXiv.2412.14737},
	year = {2024},
	note = {arXiv:2412.14737},
	title = {On {Verbalized} {Confidence} {Scores} for {LLMs}},
}

@inproceedings{xu2025mirror,
	author = {Xu, Changye and Wen, Bingbing and Han, Bohan and Wolfe, Robert and Wang, Lucy Lu and Howe, Bill},
	booktitle = {Findings of the {Association} for {Computational} {Linguistics}: ACL 2025},
	doi = {10.18653/v1/2025.findings-acl.1316},
	year = {2025},
	note = {arXiv:2506.00582},
	title = {Do {Language} {Models} {Mirror} {Human} {Confidence}?},
}

@inbook{hosmer2013assessing,
	author = {Hosmer, David W. and Lemeshow, Stanley and Sturdivant, Rodney X.},
	booktitle = {Applied {Logistic} {Regression}},
	doi = {10.1002/9781118548387.ch5},
	isbn = {978-1-118-54838-7},
	year = {2013},
	pages = {153--225},
	publisher = {John Wiley \& Sons},
	title = {Assessing the {Fit} of the {Model}},
}

@inproceedings{guo2017calibration,
	author = {Guo, Chuan and Pleiss, Geoff and Sun, Yu and Weinberger, Kilian Q.},
	series = {PMLR},
	booktitle = {Proceedings of the 34th {International} {Conference} on {Machine} {Learning} ({ICML})},
	year = {2017},
	pages = {1321--1330},
	title = {On {Calibration} of {Modern} {Neural} {Networks}},
	volume = {70},
}

@article{cohen1960coefficient,
	author = {Cohen, Jacob},
	journal = {Educational and Psychological Measurement},
	doi = {10.1177/001316446002000104},
	number = {1},
	year = {1960},
	pages = {37--46},
	title = {A {Coefficient} of {Agreement} for {Nominal} {Scales}},
	volume = {20},
}

@article{gupta2024cascades,
	author = {Gupta, Neha and Narasimhan, Harikrishna and Jitkrittum, Wittawat and Rawat, Ankit Singh and Menon, Aditya Krishna and Kumar, Sanjiv},
	doi = {10.48550/arXiv.2404.10136},
	year = {2024},
	note = {arXiv:2404.10136},
	title = {Language {Model} {Cascades}: Token-{Level} {Uncertainty} and {Beyond}},
}

@article{zellinger2024hcma,
	author = {Zellinger, Michael J. and Thomson, Matt},
	doi = {10.48550/arXiv.2410.02173},
	year = {2024},
	note = {arXiv:2410.02173},
	title = {Efficiently {Deploying} {LLMs} with {Controlled} {Risk}},
}

@article{degroot1983comparison,
	author = {DeGroot, Morris H. and Fienberg, Stephen E.},
	journal = {The Statistician},
	doi = {10.2307/2987588},
	number = {1/2},
	year = {1983},
	pages = {12--22},
	title = {The {Comparison} and {Evaluation} of {Forecasters}},
	volume = {32},
}

\end{document}